\documentclass[3p]{elsarticle}

\usepackage{amssymb,amsmath}
\usepackage{bm}
\usepackage{graphicx}
\usepackage{tabularx}
\usepackage{comment}
\usepackage{fancyvrb}

\includecomment{pdffigure}

\DefineVerbatimEnvironment{code}{Verbatim}{fontsize=\small}
\newcommand{\openone}[0]{\leavevmode\hbox{\small1\normalsize\kern-.33em1}}

\biboptions{sort&compress}

\begin{document}

\begin{frontmatter}


\title{Simulating electron energy loss spectroscopy with the MNPBEM toolbox}

\author{Ulrich Hohenester}
\ead{ulrich.hohenester@uni-graz.at}
\ead[url]{http://physik.uni-graz.at/~uxh}
\address{Institut f\"ur Physik. Karl--Franzens--Universit\"at Graz, 
  Universit\"atsplatz 5, 8010 Graz, Austria}



\begin{abstract}

Within the \texttt{MNPBEM} toolbox, we show how to simulate electron energy loss spectroscopy (EELS) of plasmonic nanoparticles using a boundary element method approach.  The methodology underlying our approach closely follows the concepts developed by Garc\'\i a de Abajo and coworkers [for a review see Rev. Mod. Phys. 82, 209 (2010)].  We introduce two classes \texttt{eelsret} and \texttt{eelsstat} that allow in combination with our recently developed \texttt{MNPBEM} toolbox for a simple, robust, and efficient computation of EEL spectra and maps.  The classes are accompanied by a number of demo programs for EELS simulation of metallic nanospheres, nanodisks, and nanotriangles, and for electron trajectories passing by or penetrating through the metallic nanoparticles.   We also discuss how to compute electric fields induced by the electron beam and cathodoluminescence.

\end{abstract}


\end{frontmatter}

\section*{Program summary}

\noindent{\sl Program title:} \texttt{MNPBEM} toolbox supplemented by a collection of demo files\\
\noindent{\sl Programming language:} Matlab 7.11.0 (R2010b)\\
\noindent{\sl Computer:} Any which supports Matlab 7.11.0 (R2010b)\\
\noindent{\sl Operating system:} Any which supports Matlab 7.11.0 (R2010b)\\
\noindent{\sl RAM required to execute with typical data:} $\ge 1$ GByte\\
\noindent{\sl Has the code been vectorised or parallelized?:} yes\\
\noindent{\sl Keywords:} Plasmonics, electron energy loss spectroscopy, boundary element method\\
\noindent{\sl CPC Library Classification:} Optics\\
\noindent{\sl External routines/libraries used:} \texttt{MESH2D} available at \texttt{www.mathworks.com}\\
\noindent{\sl Nature of problem:} Simulation of electron energy loss spectroscopy (EELS) for plasmonic nanoparticles\\
\noindent{\sl Solution method:} Boundary element method using electromagnetic potentials\\
\noindent{\sl Running time:} Depending on surface discretization between seconds and hours\\

\section{Introduction}\label{sec:intro}

Plasmonics has emerged as an ideal tool for light confinement at the nanoscale~\cite{maier:07,atwater:07,schuller:10,stockman:11}.  This is achieved through light excitation of coherent charge oscillations at the interface between metallic nanoparticles and a surrounding medium, the so-called \textit{surface plasmons}, which come together with strongly localized, evanescent fields.  While the driving force behind plasmonics is downscaling of optics to the nanoscale, conventional optics cannot be used for mapping of plasmonic fields because to the Abbe diffraction limit of light.  To overcome this limit, various experimental techniques, such as scanning near field microscopy or scanning tunneling spectroscopy~\cite{novotny:06}, have been employed.

In recent years, electron energy loss spectroscopy (EELS) has become an extremely powerful experimental device for the minute spatial and spectral investigation of plasmonic fields~\cite{garcia:10}.  In EELS, electrons with a high kinetic energy pass by or penetrate through a metallic nanoparticle, excite particle plasmons, and lose part of their kinetic energy.  By monitoring this energy loss as a function of electron beam position, one obtains a detailed map about the localized plasmonic fields~\cite{bosman:07,nelayah:07}.  This technique has been extensively used in recent years to map out the plasmon modes of nanotriangles~\cite{nelayah:07,nelayah:09,schaffer:10}, nanorods~\cite{bosman:07,schaffer.prb:09,nicoletti:11,rossouw:13}, nanodisks~\cite{schmidt:12}, nanocubes~\cite{mazzucco:12}, nanoholes~\cite{sigle:09}, and coupled nanoparticles~\cite{chu:09,ngom:09,koh:09,koh:11} (see also Refs.~\cite{garcia:08,hohenester.prl:09,boudarham:12,hoerl.prl:13} for the interpretation of EELS maps).

Simulation of EEL spectra and maps has primarily been performed with the discrete-dipole approximation~\cite{geuquet:10,bigelow:12} and the boundary element method (BEM) approach~\cite{myroshnychenko:08,garcia:10}.  Within the latter scheme, the boundary of the metallic nanoparticle becomes discretized by boundary elements, and Maxwell's equations are solved by attaching artificial surface charges and currents to these elements which are chosen such that the proper boundary conditions are fulfilled~\cite{garcia:02,hohenester.cpc:12}.  The methodology for EELS simulations within the BEM approach has been developed in Refs.~\cite{garcia:97,garcia:02,garcia:10}, and has been successfully employed in comparison with experimental EELS data~\cite{nelayah:07,schaffer.prb:09,schmidt:12,mazzucco:12}.

\subsection{Purpose of EELS software and its relation to the MNPBEM toolbox}

\begin{figure}
\centerline{\includegraphics[width=0.7\columnwidth]{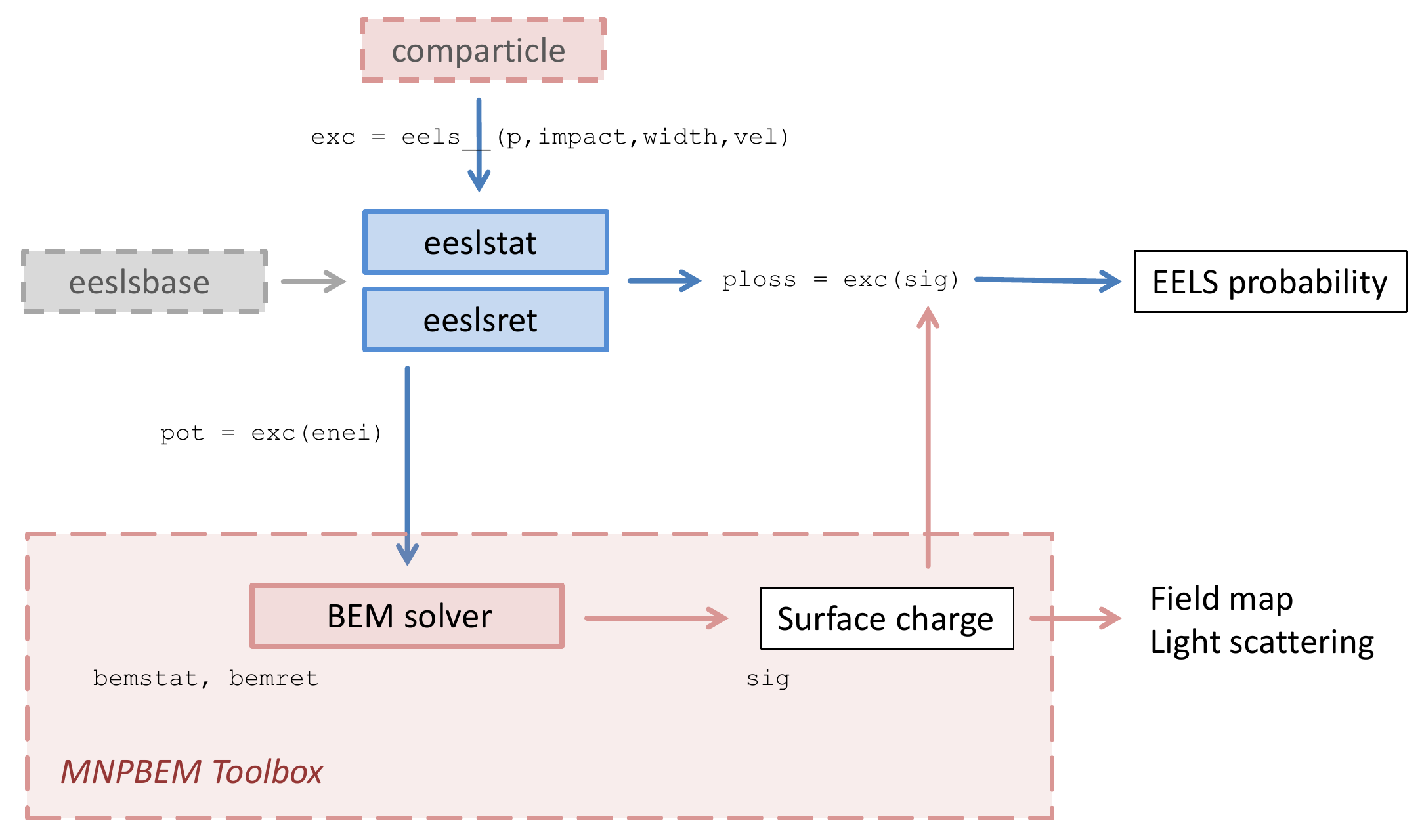}}
\caption{Overview of the EELS software and its relation to \texttt{MNPBEM} toolbox.  The classes \texttt{eelsstat} and \texttt{eelsret} for the simulation of electron energy loss of plasmonic nanoparticles, within the quasistatic limit or for the full Maxwell equations, are initialized with a \texttt{comparticle} object~\cite{hohenester.cpc:12}, which stores the particle boundaries and the dielectric materials, together with the impact parameters of the electron beam.  The EELS classes provide the external potentials, which can be used by the \texttt{MNPBEM} toolbox to compute the surface charges \texttt{sig}, which, in turn, allow to determine electron energy loss probabilities as well as field maps or scattered light (cathodoluminescence).}\label{fig:flowchart}
\end{figure}

The purpose of the EELS software described in this paper is to allow for a simple and efficient computation of electron energy loss spectroscopy of plasmonic nanoparticles and other nanophotonic structures.  The software consists of two classes \texttt{eelsret} and \texttt{eelsstat} devoted to the simulation of EEL spectroscopy and mapping of plasmonic nanoparticles, see Fig.~\ref{fig:flowchart}, which can be used in combination with the \texttt{MNPBEM} toolbox~\cite{hohenester.cpc:12} that provides a generic simulation platform for the solution of Maxwell's equations.  Our implementation for EELS simulation of plasmonic nanoparticles relies on a BEM approach~\cite{garcia:02,garcia:10} that has been successfully employed in various studies~\cite{schaffer.prb:09,hohenester.prl:09,schmidt:12,hoerl.prl:13}.  A typical simulation scenario consists of the following steps.

\begin{enumerate}

\item First, one sets up the particle boundaries and the dielectric environment within which the nanoparticle is embedded.  This step has been described in detail in our previous \texttt{MNPBEM} paper~\cite{hohenester.cpc:12}.

\item We next initialize an \texttt{eelsret} or \texttt{eelsstat} object which defines the electron beam.  This object stores the beam positions and the electron velocity.  For a given electron loss energy, it then returns the external scalar and vector potentials $\phi_{\rm ext}$ and $\bm A_{\rm ext}$ which can be used for the solution of the BEM equations~\cite{garcia:02}.

\item For given $\phi_{\rm ext}$ and $\bm A_{\rm ext}$, we solve the full Maxwell equations or its quasistatic limit, using the classes \texttt{bemret} or \texttt{bemstat} of the \texttt{MNPBEM} toolbox~\cite{hohenester.cpc:12}.  The solutions are provided by the surface charge and current distributions $\sigma$ and $\bm h$, which allow to compute the potentials and fields at the particle boundary and everywhere else~\cite{garcia:02} (using the Green function of the Helmholtz equation).

\item Finally, we use $\sigma$ and $\bm h$ to compute the electron energy loss probabilities, which can be directly compared with experimental EEL data.

\end{enumerate}

Rather than providing the additional classes separately, we have embedded them in a new version of the \texttt{MNPBEM} toolbox which supersedes the previous version~\cite{hohenester.cpc:12}.  The main reason for this policy is that also the Mie classes \texttt{mieret} and \texttt{miestat} had to be modified, which allow a comparison with analytic Mie results and can be used for testing.  The new version of the toolbox also corrects a few minor bugs and inconsistencies.  However, we expect that all simulation programs that performed with the old version should also work with the new version.

We have organized this paper as follows.  In Sec.~\ref{sec:start} we discuss how to install the toolbox and give a few examples demonstrating the performance of EELS simulations.  The methodology underlying our approach as well as a few implementation details are presented in Sec.~\ref{sec:theory}.  Finally, in Sec.~\ref{sec:results} we present results of our EELS simulations and provide a detailed toolbox description.

\section{Getting started}\label{sec:start}

\subsection{Installation of the toolbox}

To install the toolbox, one must simply add the path of the main directory \texttt{mnpbemdir} of the \texttt{MNPBEM} toolbox as well as the paths of all subdirectories to the Matlab search path.  This can be done, for instance, through
\begin{code}
addpath(genpath(mnpbemdir));
\end{code}
To set up the help pages, one must once change to the main directory of the \texttt{MNPBEM} toolbox and run the program \texttt{makemnpbemhelp}
\begin{code}
>> cd mnpbemdir;
>> makemnpbemhelp;
\end{code}
Once this is done, the help pages, which provide detailed information about the toolbox, are available in the Matlab help browser.  Note that one may have to call \texttt{Start > Desktop Tools > View Start Button Configuration > Refresh} to make the help accessible.  Under Matlab 2012 the help pages can be found on the start page of the help browser under \textit{Supplemental Software}.  The toolbox is almost identical to our previously published version~\cite{hohenester.cpc:12}.  The only major difference concerns the inclusion of EELS simulations, which will be described in more detail in this paper.

\subsection{Brief overview of the EELS software}

The \texttt{MNPBEM} toolbox comes together with a directory \texttt{demoeels} containing several demo files.  To get a first impression, we recommend to work through these demo files.  By changing to the demo directory and typing 

\begin{figure}
\centerline{\includegraphics[width=0.75\columnwidth]{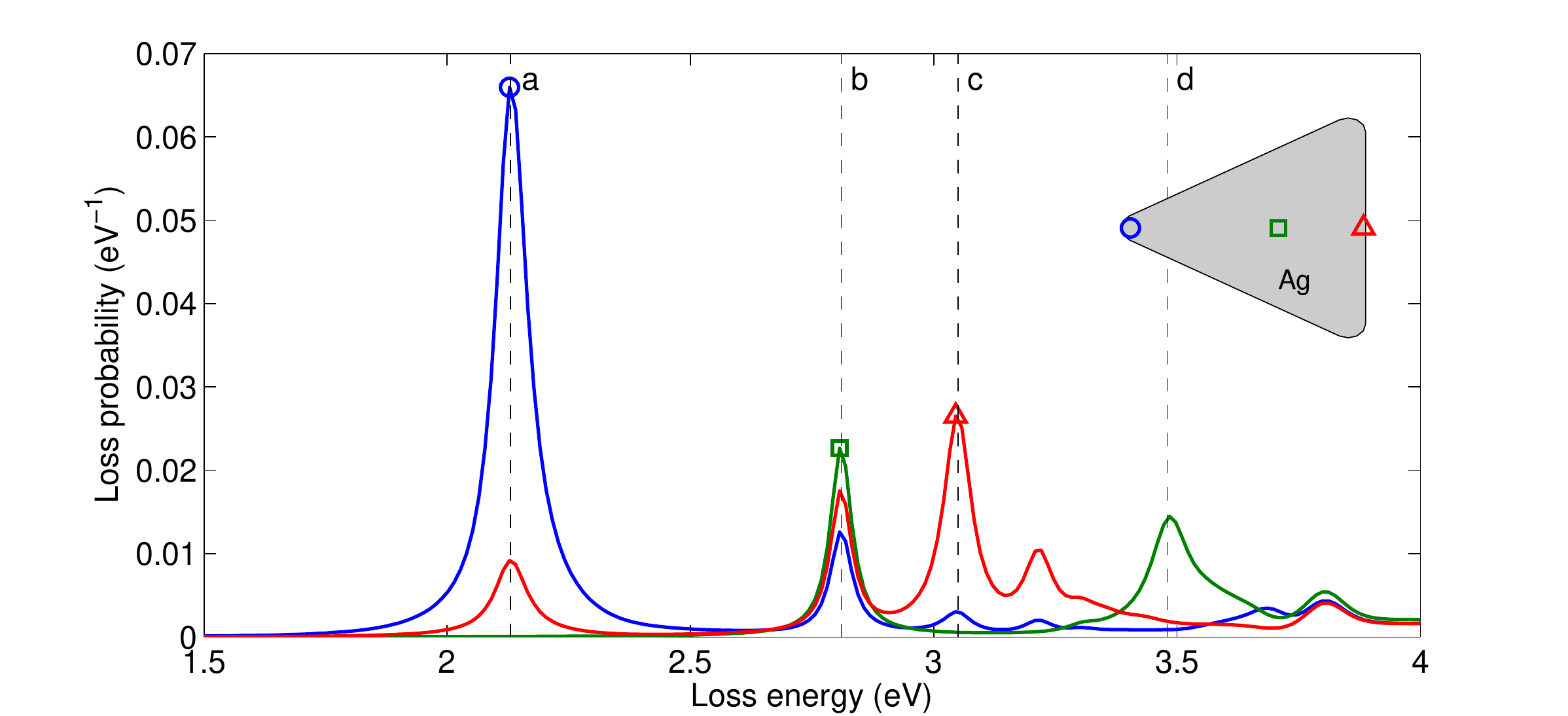}}
\caption{EEL spectra at three different positions of the electron beam, indicated in the inset, as computed with the MNPBEM toolbox and the demo program \texttt{demotrianglespectrum.m} for a silver triangle with a base length of approximately 80 nm and a height of 10 nm.  In our simulations we use dielectric functions extracted from optical experiments~\cite{johnson:72}.  The dashed lines report the energetic positions of the plasmon resonances where the spatial EELS maps of Fig.~\ref{fig:trianglemap} are computed.}\label{fig:trianglespectrum}
\end{figure}

\begin{figure}
\centerline{\includegraphics[width=0.6\columnwidth]{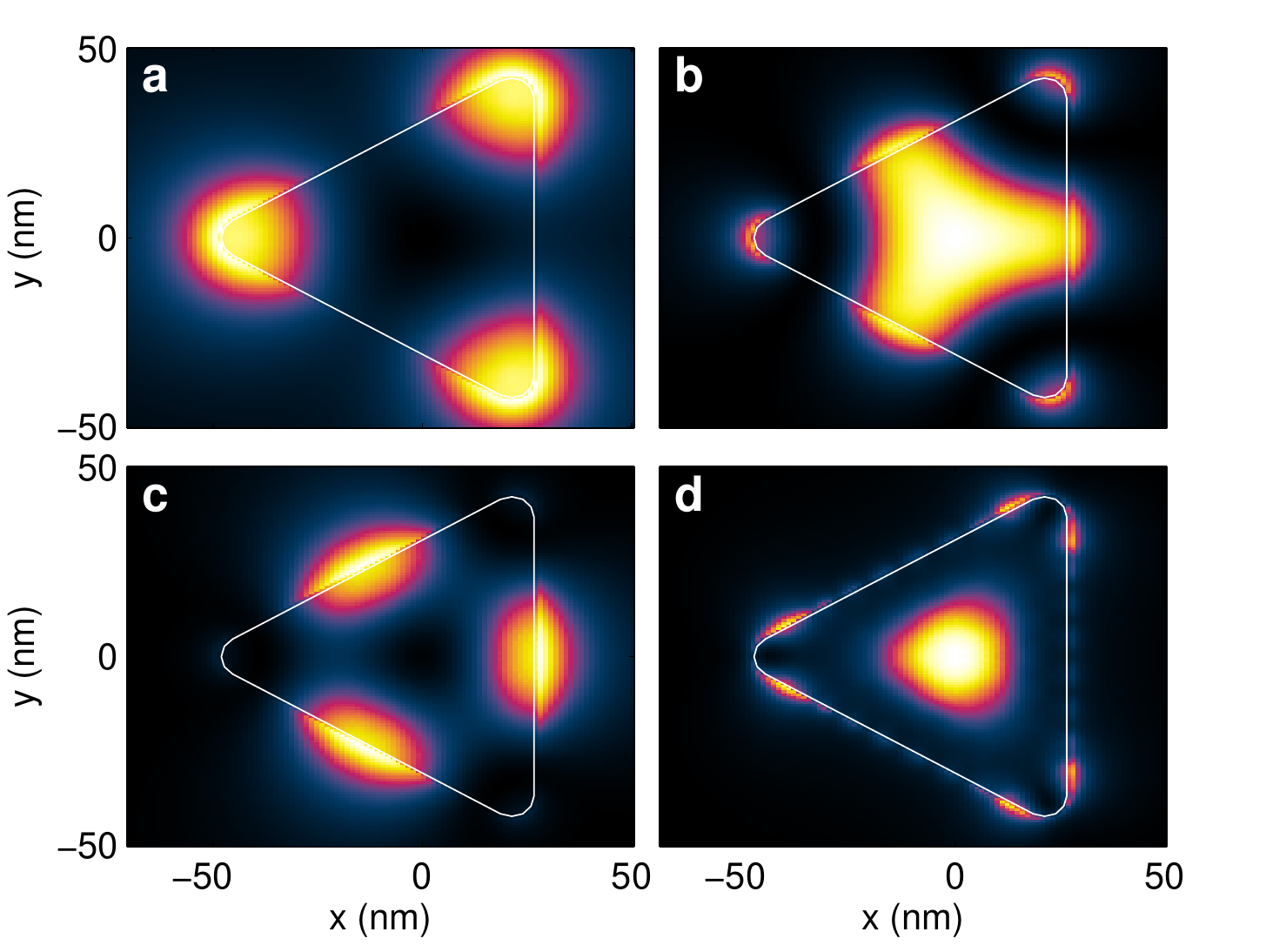}}
\caption{Spatial EELS maps for same nanotriangle as investigated in Fig.~\ref{fig:trianglespectrum} at the plasmon resonances indicated by dashed lines.  The maps have been computed with the demo program \texttt{demotrianglemap.m}, with color axes scaled to the maxima of the respective maps (see Fig.~\ref{fig:diskmap} for color bar).}\label{fig:trianglemap}
\end{figure}

\begin{code}
>> demotrianglespectrum
\end{code}
at the Matlab prompt, a simulation is performed where the EEL spectra are computed for a triangular silver nanoparticle.  The run time is reported in Table~\ref{table:examples}, and the simulation results are shown in Fig.~\ref{fig:trianglespectrum}.  One observes a number of peaks associated with the different plasmon modes of the nanoparticle.  Note that through \verb!plot(p,'EdgeColor','b')! one can plot the nanoparticle boundary.  By running next \texttt{demotrianglemap.m} we obtain the spatial EELS maps at the plasmon resonance energies indicated by dashed lines in Fig.~\ref{fig:trianglespectrum}.  Figure~\ref{fig:trianglemap}(a) reports the map for the degenerate dipolar modes, whereas panels (b--d) show the EELS maps for higher excited plasmon modes (see Ref.~\cite{nelayah:07} for experimental results).

\begin{table}[b]
\caption{Demo programs for EELS simulations provided by the \texttt{MNPBEM} toolbox.  We list the names of the programs, typical runtimes, and give brief explanations.  The programs were tested on a standard PC (Intel i7--2600 CPU, 3.40 GHz, 8 GB RAM).}\label{table:examples}
{\small
\begin{tabularx}{\columnwidth}{lrX}
\hline\hline
Demo program & Runtime & Description \\
\hline
\texttt{demomie.m} & 26.3 sec & Comparison of BEM simulations with analytic Mie results\\
\texttt{demomiestat.m} & 14.7 sec & Same as \texttt{demomie.m} but for quasistatic limit\\
\texttt{demodiskspectrum.m} & 29.4 sec & EEL spectra for nanodisk at selected beam positions\\
\texttt{demodiskspectrumstat.m} & 14.6 sec & Same as \texttt{demodiskspectrum.m} but for quasistatic limit\\
\texttt{demodiskmap.m} & 105.6 sec & Spatial EELS maps for nanodisk at selected loss energies\\
\texttt{demotrianglespectrum.m} & 173.5 sec & EEL spectra for nanotriangle at selected beam positions\\
\texttt{demotrianglemap.m} &  104.6 sec & Spatial EELS maps for nanotriangle at selected loss energies\\
\hline
\hline
\end{tabularx}}
\end{table}

\subsection{A first example}

In the following we briefly discuss the demo file \texttt{demotrianglespectrum.m} (see Sec.~\ref{sec:results} for a detailed description of the software).  We first set up a \texttt{comparticle} object \texttt{p} for the nanotriangle, which stores the particle boundary and the dielectric materials, as well as a BEM solver \texttt{bem} for the solution of Maxwell's equations.  These steps have been described in some length in a previous paper~\cite{hohenester.cpc:12}.  Additional information is also provided by the help pages.  To set up the EELS simulation, we need the impact parameters of the electron beam, a broadening parameter for the triangle integration (see Sec.~\ref{sec:penetrate} for details), and the electron velocity in units of the speed of light.  Initialization is done through
\begin{code}
b = [ - 45, 0; 0, 0; 26, 0 ];       %  impact parameters (triangle corner, middle, edge)
vel = 0.7;                          %  electron velocity in units of speed of light
width = 0.2;                        %  broadening of electron beam
exc = eelsret( p, b, width, vel );  %  initialize EELS excitation
\end{code}
The \texttt{exc} object returns through \verb!exc(enei)! the external potentials for a given loss energy, which allow for the solution of the BEM equations by means of artificial surface charges and currents \texttt{sig}~\cite{garcia:02,hohenester.cpc:12}.  From \texttt{sig} we can obtain the surface and bulk loss probabilities for the electron.
\begin{code}
sig = bem \ exc( enei );             %  surface charges and currents 
[ psurf, pbulk ] = exc.loss( sig );  %  surface and bulk losses
\end{code}
Finally, the EEL spectrum can be computed by performing a loop over loss energies, and EEL maps can be obtained by providing a rectangular grid of impact parameters.  A more detailed description of the EELS classes will be given in Sec.~\ref{sec:results}.

\section{Theory}\label{sec:theory}

\subsection{Boundary element method (BEM)}

For the sake of completeness, we start by briefly summarizing the main concepts of the BEM approach for the solution of Maxwell's equation (see Refs.~\cite{garcia:02,garcia:10,hohenester.cpc:12} for a more detailed discussion).  We consider dielectric nanoparticles, described through local and isotropic dielectric functions $\varepsilon_j(\omega)$, which are separated by sharp boundaries $\partial V_j$.  Throughout, we set the magnetic permeability $\mu=1$ and consider Maxwell's equations in frequency space $\omega$ \cite{jackson:99}.  In accordance to Refs.~\cite{garcia:02,garcia:10,hohenester.cpc:12}, we adopt a Gaussian unit system.

The basic ingredients of the BEM approach are the scalar and vector potentials $\phi(\bm r)$ and $\bm A(\bm r)$, which are related to the electromagnetic fields via 
\begin{equation}
  \bm E=ik\bm A-\nabla\phi\,,\quad\bm B=\nabla\times\bm A\,.
\end{equation}
Here $k=\omega/c$ and $c$ are the wavenumber and speed of light in vacuum, respectively.  The potentials are connected through the Lorentz gauge condition $\nabla\cdot\bm A=ik\varepsilon\phi$.  Within each medium, we introduce the Green function for the Helmholtz equation defined through
\begin{equation}\label{eq:greenret}
  \left(\nabla^2+k_j^2\right)G_j(\bm r,\bm r')=-4\pi\delta(\bm r-\bm r')\,,\quad
  G_j(\bm r,\bm r')=\frac{e^{ik_j|\bm r-\bm r'|}}{|\bm r-\bm r'|}\,,
\end{equation}
with $k_j=\sqrt{\varepsilon_j}k$ being the wavenumber in the medium $\bm r\in V_j$.  For an inhomogeneous dielectric environment, we then write down the solutions of Maxwell's equations in the \textit{ad-hoc} form \cite{garcia:02,garcia:10}
\begin{subequations}\label{eq:adhoc}
\begin{eqnarray}
  \phi(\bm r)&=&\phi_{\rm ext}(\bm r)+
    \oint_{V_j} G_j(\bm r,\bm s)\sigma_j(\bm s)\,da\label{eq:adhocphi}\\
  \bm A(\bm r)&=&\bm A_{\rm ext}(\bm r)+
    \oint_{V_j} G_j(\bm r,\bm s)\bm h_j(\bm s)\,da\,, \label{eq:adhoca}
\end{eqnarray}
\end{subequations}
where $\phi_{\rm ext}$ and $\bm A_{\rm ext}$ are the scalar and vector potentials characterizing the external perturbation. Owing to Eq.~\eqref{eq:greenret}, these expressions fulfill the Helmholtz equations everywhere except at the particle boundaries.  $\sigma_j$ and $\bm h_j$ are surface charge and current distributions, which are chosen such that the boundary conditions of Maxwell's equations at the interfaces between regions of different permittivies $\varepsilon_j$ hold.  This leads to a number of integral equations.  Upon discretization of the particle boundaries into boundary elements, one obtains a set of linear equations that can be inverted, thus providing the solutions of Maxwell's equation in terms of surface charge and current distributions $\sigma_j$ and $\bm h_j$.  Through Eqs.~(\ref{eq:adhoc}a,b) one can compute the potentials everywhere else.  For further details about the working equations of the BEM approach the reader is referred to Refs.~\cite{garcia:02,garcia:10,hohenester.cpc:12}.

\subsection{Electron energy loss spectroscopy (EELS)}

In the following we consider the situation where an electron passes by or penetrates through a metallic nanoparticles, and loses energy by exciting particle plasmons.  We assume that the electron kinetic energy is much higher than the plasmon energies (for typical electron microscopes operating with electron energies of several hundreds of keV this assumption is certainly fulfilled).  We can thus discard in the electron trajectory the small change of velocity due to plasmon excitation, and describe the loss process in lowest order perturbation theory.  We emphasize that our approach is correspondingly not suited for low electron energies or thick samples.

For an electron trajectory $\bm r(t)=\bm r_0+\bm vt$, with $\bm v=v\hat{\bm z}$, the electron charge distribution reads
\begin{equation}\label{eq:elecharge}
  \rho(\bm r,\omega)=-e\int e^{i\omega t}\delta(\bm r-\bm r_0-\bm vt)\,dt
  =-\frac ev\delta(\bm R-\bm R_0)e^{iq(z-z_0)}\,.
\end{equation}
Here $-e$ and $v$ are the charge and velocity of the electron, respectively, $\bm R_0$ is the impact parameter in the $xy$-plane, and $q=\omega/v$ is a wavenumber.  The potentials associated with the charge distribution of Eq.~\eqref{eq:elecharge} can be computed in infinite space analytically (Li\'enard-Wiechert potentials \cite{jackson:99}), and we obtain \cite{garcia:02,garcia:10}
\begin{equation}\label{eq:lienard}
  \phi_{\rm ext}(\bm r)=-\frac 2{v\varepsilon_j}K_0\left(\frac{q|\bm R-\bm R_0|}{\gamma_j}\right)
  e^{iq(z-z_0)}\,,\quad 
  \bm A_{\rm ext}(\bm r)=\varepsilon_j\frac{\bm v}c\phi_{\rm ext}(\bm r)\,.
\end{equation}
$K_0$ is the modified Bessel function of order zero, and $\gamma_j=(1-\varepsilon_j v^2/c^2)^{-\frac 12}$.  Within our BEM approach, we can directly insert the expressions of Eq.~\eqref{eq:lienard} for the unbounded medium into Eq.~\eqref{eq:adhoc} since the calculated surface charge and current distributions $\sigma_j$ and $\bm h_j$ will automatically guarantee that the proper boundary conditions at the interfaces are fulfilled~\cite{garcia:02}.  

We next turn to the calculation of the electron energy loss.  Ignoring the small change of the electron velocity caused by the interaction with the plasmonic nanoparticle, the energy loss can be computed from the work performed by the electron against the induced field~\cite{ritchie:57,garcia:02,garcia:10}
\begin{equation}\label{eq:elework}
  \Delta E=e\int \bm v\cdot\bm E_{\rm ind}[\bm r(t),t]\,dt=\int_0^\infty \hbar\omega
  \Gamma_{\rm EELS}(\bm R,\omega)\,d\omega\,,
\end{equation}
with the loss probability, given per unit of transferred energy,
\begin{equation}\label{eq:eleloss}
  \Gamma_{\rm EELS}(\bm R,\omega)=\frac e{\pi\hbar\omega}\int
  \Re e\bigl\{e^{-i\omega t}\bm v\cdot\bm E_{\rm ind}[\bm r(t),\omega]\bigr\}\,dt+
  \Gamma_{\rm bulk}(\omega)\,.
\end{equation}
Note that Eq.~\eqref{eq:eleloss} is a classical expression, where $\hbar$ has been introduced only to relate energy and frequency.  $\bm E_{\rm ind}$ is the induced electric field, which can be computed from the potentials originating from the surface charge and current distributions $\sigma_j$ and $\bm h_j$ alone.  $\Gamma_{\rm bulk}$ is the bulk loss probability for electron propagation inside a lossy medium, see Eq.~(18) of Ref.~\cite{garcia:10}.  Within the quasistatic approximtion, it is proportional to the loss function $\propto\Im m[-1/\varepsilon(\omega)]$ and the propagation length inside the medium.  Expressions similar to Eq.~\eqref{eq:eleloss} but derived within a full quantum approach, based on the Born approximation, can be found in Ref.~\cite{garcia:10}.

Quite generally, $\Gamma_{\rm EELS}$ can be computed by calculating the induced electric field along the electron trajectory and evaluating the expression given in Eq.~\eqref{eq:eleloss}.  In what follows, we describe a computationally more efficient scheme.  Insertion of the induced potentials of Eq.~\eqref{eq:adhoc} into the energy loss expression of Eq.~\eqref{eq:eleloss} yields
\begin{equation}\label{eq:eleloss2}
  \Gamma_{\rm EELS}(\bm R,\omega)=\frac e{\pi\hbar\omega}\sum_j\int_{z_j^0}^{z_j^1}\Re e\left[
  e^{-iqz}\oint_{\partial V_j}\bm v\cdot\Bigl\{
  ikG_j(\bm r-\bm s)\bm h_j(\bm s)-\nabla G_j(\bm r-\bm s)\sigma_j(\bm s)\Bigr\}
  \,da\right]dz\,.
\end{equation}
Here $z_j^0$ and $z_j^1$ are the entrance and exit points of the electron beam in a given medium, and $\bm r=\bm R+z\hat{\bm z}$ parameterizes the electron trajectory.  We next introduce a potential-like term $\varphi_j(\bm s)=\int_{z_j^0}^{z_j^1} e^{-iqz} G_j(\bm r-\bm s)\,dz$, associated with the electron propagation inside a given medium.  Performing integration by parts, we can rewrite the second term in parentheses of Eq.~\eqref{eq:eleloss2} as
\begin{displaymath}
  \int_{z_j^0}^{z_j^1} e^{-iqz} \frac{\partial G_j(\bm r-\bm s)}{\partial z}\,dz=
  e^{-iqz}G_j(\bm r-\bm s)\Bigl|_{z_j^0}^{z_j^1}+iq\,e^{-iqz}\varphi_j(\bm s)\,.
\end{displaymath}
The first term on the right-hand side gives, upon insertion into Eq.~\eqref{eq:eleloss2}, $e^{-iqz}\oint G_j(\bm r-\bm s)\sigma_j(\bm s)\,da\,\bigl|_{z_j^0}^{z_j^1}$.  The integral expression precisely corresponds to the scalar potential at the crossing points of the trajectory with the particle boundary.  As the potential is continuous across the boundaries, the contributions of all crossing points sum up to zero.  Thus, we arrive at our final result
\begin{equation}\label{eq:gammaeels}
  \Gamma_{\rm EELS}(\bm R,\omega)=-\frac e{\pi\hbar\omega}\sum_j\Im m\left[
  \oint_{\partial V_j}\varphi_j(\bm s)\Bigl\{k\bm v\cdot\bm h_j(\bm s)-q v\sigma_j(\bm s)\Bigr\}da
  \right]\,.
\end{equation}
In comparison to Eq.~\eqref{eq:eleloss2}, this expression has the advantage that the integration is only performed over the particle boundary, where the surface charge and current distributions $\sigma_j$ and $\bm h_j$ are readily available, and we don't have to compute the induced electric field along the electron trajectory.

\subsection{Refined integration over boundary elements}\label{sec:refine}

In the calculation of the external potential, Eq.~\eqref{eq:lienard}, and the EELS probability of Eq.~\eqref{eq:gammaeels} the points where the electron trajectory crosses the boundary have to be treated with care.  For small distances, the potential scales with $K_0(\rho)\sim -\log\rho$.  When integrating this expression within our BEM approach over a small area, we find in polar coordinates that $\int K_0(\rho)\rho\,d\rho$ remains finite.  The same is true for the surface derivative of the potential.

In a computational approach it is somewhat tedious to perform such integration properly, in particular for crossing points that are located close to the edges or corners of boundary elements.  For this reason, we suggest a slightly different approach.  The main idea is to replace the Delta-like transversal trajectory profile $\delta(\rho)$ by a smoothened distribution.  The potential at the transverse position $\bm R=(x,0)$ then reads
\begin{equation}\label{eq:profile}
  \phi(x,0)={\rm const}\times \int_0^\infty\rho d\rho\int_0^{2\pi}d\theta\,
  K_0\left(\lambda\sqrt{(x-\rho\cos\theta)^2+(\rho\sin\theta)^2}\right)
  \left[\frac 1\pi\frac{\rho_0^2}{(\rho^2+\rho_0^2)^2}\right]\,.
\end{equation}
Here $\lambda=q/\gamma$ [see Eq.~\eqref{eq:lienard}] and the term in brackets is our smoothing function, with $\rho_0$ being a parameter that determines the transversal extension.  For small arguments, we can expand $K_0(\lambda r)\approx \gamma-\log(\lambda r/2)$, where $\gamma$ is the Euler constant, and perform all integrations analytically to obtain $\gamma-\log(\lambda\sqrt{x^2+\rho_0^2}/2)$.  This suggests replacing the potential of Eq.~\eqref{eq:lienard} by the smoothened function
\begin{equation}\label{eq:phismooth}
  \phi_{\rm ext}(\bm r)=-\frac 2{v\varepsilon_j}K_0\left(\frac{q\tilde R}
  {\gamma_j}\right)e^{iq(z-z_0)}\,,\quad \tilde R=\sqrt{|\bm R-\bm R_0|^2+\rho_0^2}\,.
\end{equation}
For large arguments this expression coincides with the Li\'enard-Wiechert potential, but remains finite for small arguments.  A corresponding smoothening is also performed in the potential-like function $\varphi$ of Eq.~\eqref{eq:gammaeels}.

\subsection{Quasistatic approach}\label{sec:quasistatic}

In the quasistaic approximation one assumes that the size of the nanostructure $L$ is significantly smaller than the wavelength of light, such that $kL\ll 1$.  This allows us to keep in the simulations only the scalar potential and to set in the Green function $k=0$.  We are thus left with the solution of the Laplace or Poisson equation, rather than the Helmholtz equation, but we keep the full frequency-dependence of the permittivities $\varepsilon(\omega)$.  

The calculation of EELS probabilities with the BEM approach has been described in some detail in Ref.~\cite{garcia:97}.  In the following we briefly describe the basic ingredients.  First, we compute the external potential from the solution of Poisson's equation
\begin{equation}\label{eq:lienardstat}
  \phi_{\rm ext}(\bm r,\omega)=\int\frac{\rho_{\rm ext}(\bm r',\omega)}%
  {\varepsilon(\bm r',\omega)|\bm r-\bm r'|}\, dz'\,,
\end{equation}
with $\rho_{\rm ext}$ being the charge distribution of Eq.~\eqref{eq:elecharge}.  We next compute the surface charge distribution $\sigma(\bm s,\omega)$ from the solution of the boundary integral equation, which, for a nanoparticle described by a single dielectric function $\varepsilon_1$ embedded in a background of dielectric constant $\varepsilon_2$, reads~\cite{garcia:97,hohenester.cpc:12}
\begin{equation}\label{eq:bemstat}
  \Lambda(\omega)\sigma(\bm s,\omega)=\oint\frac{\partial G(\bm s,\bm s')}{\partial n}\sigma(\bm s',\omega)\,
  da'+\frac{\partial\phi_{\rm ext}(\bm s,\omega)}{\partial n}\,,\quad
  \Lambda(\omega)=\frac{\varepsilon_2(\omega)+\varepsilon_1(\omega)}%
                       {\varepsilon_2(\omega)-\varepsilon_1(\omega)}\,.
\end{equation}
Here $G$ is the static Green function and $\frac\partial{\partial n}=\hat{\bm n}\cdot\nabla$ denotes the surface derivative, where $\hat{\bm n}$ is the outer surface normal of the boundary.  For materials consisting of more than one material, Eq.~\eqref{eq:bemstat} has to be replaced by a more general expression~\cite{hohenester.cpc:12}.  Finally, we compute the electron energy loss probability from (see also Eq.~(18) of Ref.~\cite{garcia:97})
\begin{equation}
  \Gamma_{\rm EELS}(\bm R,\omega)=-\frac{2e}{\pi\hbar v}
  \oint K_0(q|\bm R-\bm R_0|)\,\Im m\Bigl\{\sigma(\bm s,\omega)e^{iqz}\Bigr\}\,da+
  \Gamma_{\rm bulk}(\omega)\,,
\end{equation}
where $\Gamma_{\rm bulk}$ is the bulk loss probability (see Eq.~(19) of Ref.~\cite{garcia:10}).

\section{Results and detailed toolbox description}\label{sec:results}

\begin{figure}
\centerline{\includegraphics[width=0.55\columnwidth]{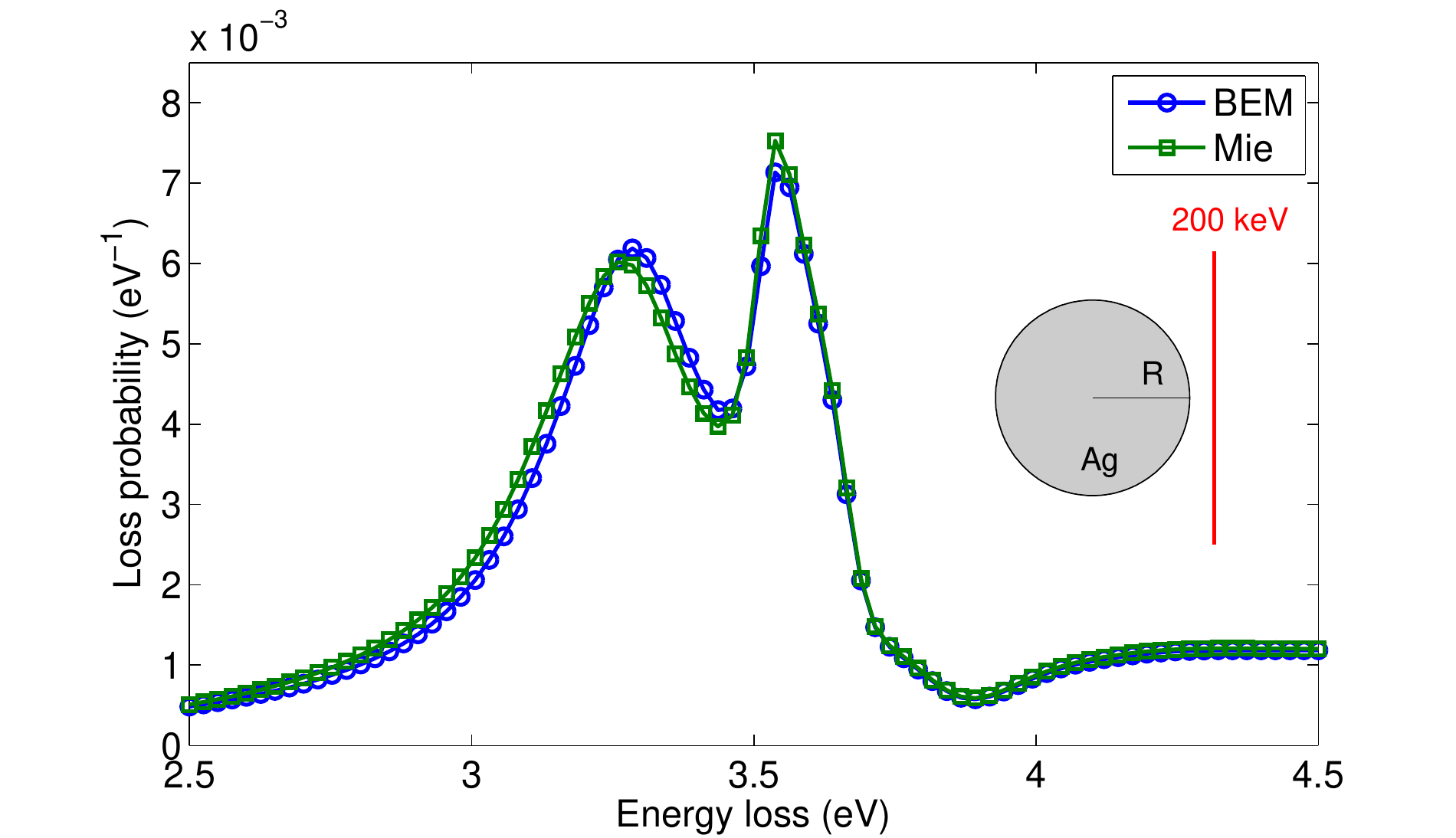}}
\caption{Loss probability for electron trajectory passing by a silver nanosphere, as shown in inset.  We compare the results of our BEM simulations with analytic results derived within Mie theory~\cite{garcia:10}.  In our simulation program \texttt{demomie.m}, the nanosphere diameter is 80 nm, the silver dielectric function is extracted from optical experiments~\cite{johnson:72}, the background dielectric constant is one, and the minimal distance between electron beam and nanosphere is 10 nm.  We assume a kinetic electron energy of 200 keV.}\label{fig:mie}
\end{figure}

We first discuss the demo file \texttt{demomie.m} that simulates the energy loss probability for an electron passing by a silver nanosphere.  Figure~\ref{fig:mie} shows results of our BEM simulations which are in good agreement with analytic Mie results~\cite{garcia:10}.  Let us briefly work through the demo program.

In the first lines we define the dielectric materials and the nanosphere (for a more detailed discussion of the \texttt{MNPBEM} toolbox see Ref.~\cite{hohenester.cpc:12}).
\begin{code}
epsm = epstable( 'silver.dat' );      %  metal dielectric function
epstab = { epsconst( 1 ), epsm };     %  table of dielectric functions
diameter = 80;                        %  sphere diameter in nanometers
%  define nanosphere and dielectric environment
p = comparticle( epstab, { trisphere( 256, diameter ) }, [ 2, 1 ], 1 );
\end{code}
We next define the excitation of the electron beam.  For the solution of the full Maxwell equations, the excitation and the calculation of the EELS probability is performed by the \texttt{eelsret} class, which is initialized through
\begin{code}
exc = eelsret( p, impact, width, vel, 'PropertyName', PropertyValue, ... )
\end{code}
Here \texttt{p} is the previously computed \texttt{comparticle} object, which stores the particle boundaries and the dielectric functions at both sides of the boundary.  \texttt{impact} is a vector \verb|[x,y]| for the impact parameter $\bm R_0=(x,y)$ of the electron beam defined in Eq.~\eqref{eq:elecharge}.  If simulations for various impact parameters are requested, as is usually the case for the simulation of EELS maps, \texttt{impact} can also be an array \verb|[x1,y1;x2,y2;...]|.  \texttt{width} is the broadening parameter $\rho_0$ of the electron beam, see Eq.~\eqref{eq:phismooth}, which will be discussed in more detail below, and \texttt{vel} is the electron velocity to be given in units of the speed of light in vacuum.  The optional pairs of property names (\verb|'cutoff'|, \verb|'rule'|, or \verb|'refine'|) and values allow to control the performance of the toolbox, as detailed in Sec.~\ref{sec:penetrate}.

In the \texttt{demomie.m} program we next set up the EELS excitation with
\begin{code}
b = 10;                               %  minimal distance from nanosphere in nanometers
vel = eelsbase.ene2vel( 200e3 );      %  electron velocity
[ width, cutoff ] = deal( 0.5, 8 );   %  width of electron beam and cutoff parameter
%  electron beam excitation for simulation of full Maxwell equations
exc = eelsret( p, [ diameter / 2 + b, 0 ], width, vel, 'cutoff', cutoff );
\end{code}
Note that \verb|eelsbase.ene2vel| allows to convert a kinetic electron energy in eV to the electron velocity in units of the speed of light in vacuum $c$.  In the above example, a kinetic energy of 200 keV corresponds to a velocity of approximately $0.7\,c$.

We next set up the solver \texttt{bemret} for the solution of the BEM equations~\cite{hohenester.cpc:12} and compute the loss probabilities of Eq.~\eqref{eq:gammaeels} for various loss energies.
\begin{code}
bem = bemret( p );               %  set up BEM solver
ene = linspace( 2.5, 4.5, 80 );  %  loss energies (in eV)
psurf = zeros( size( ene ) );    %  initialize array for surface losses

for ien = 1 : length( ene )               %  loop over energies
  sig = bem \ exc( eV2nm / ene( ien ) );  %  surface charges
  psurf( ien ) = exc.loss( sig );         %  EELS loss probabilities
end
\end{code}
Here \texttt{sig} is a \texttt{compstruct} object that stores the surface charges and current distributions $\sigma_{1,2}$ and $\bm h_{1,2}$, inside and outside the particle boundaries~\cite{garcia:02,hohenester.cpc:12}, as computed for the EELS excitation of Eq.~\eqref{eq:lienard}.  With \verb|exc.loss(sig)| we finally compute the loss probabilities according to Eq.~\eqref{eq:gammaeels}.  Note that \texttt{eV2nm} defined in \texttt{units.m} allows to convert between energies given in electronvolts and wavelengths given in nanometers, the latter being the units used by the \texttt{MNPBEM} toolbox.

\begin{figure}
\centerline{\includegraphics[height=0.3\columnwidth]{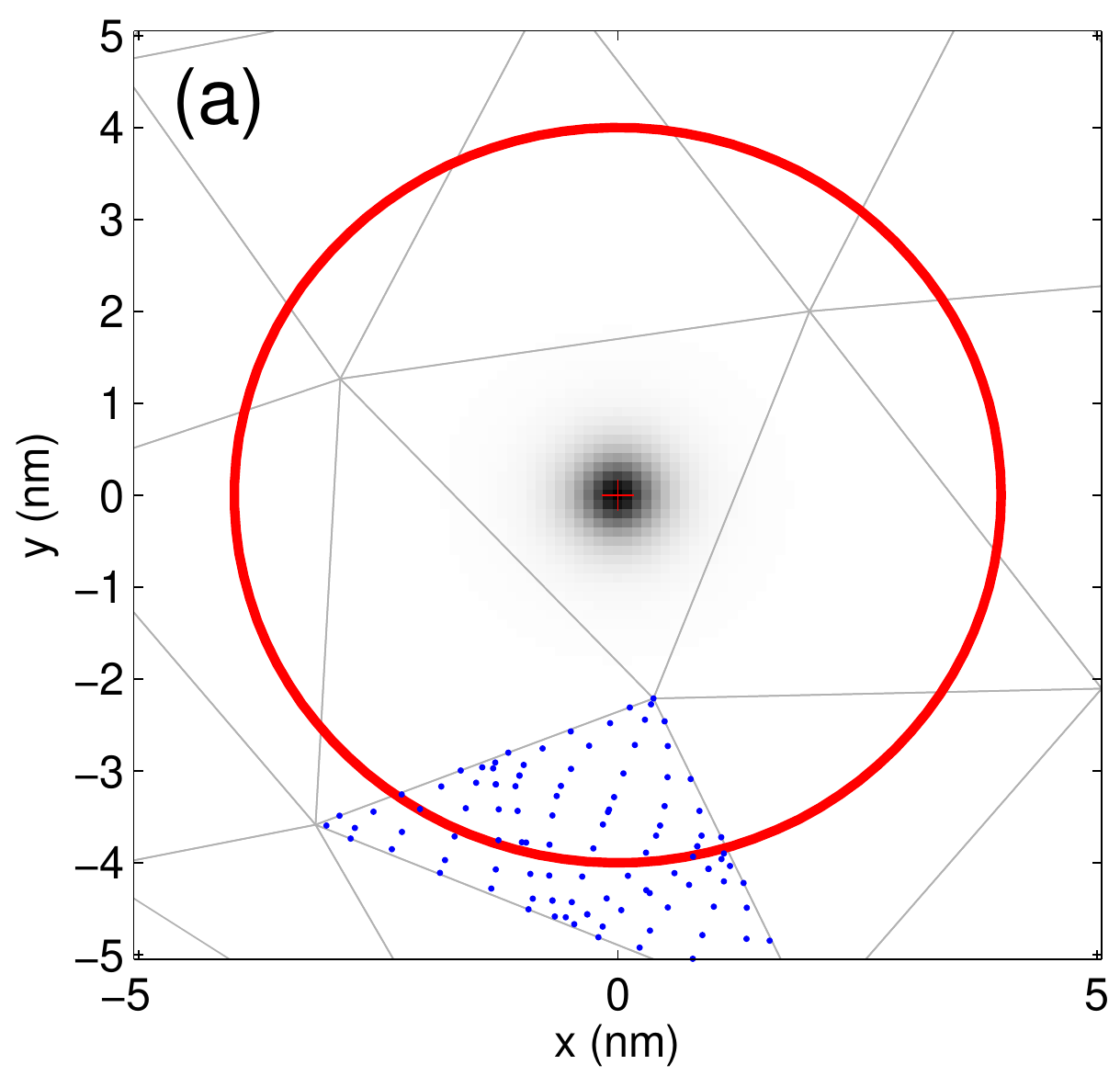}\quad
            \includegraphics[height=0.3\columnwidth]{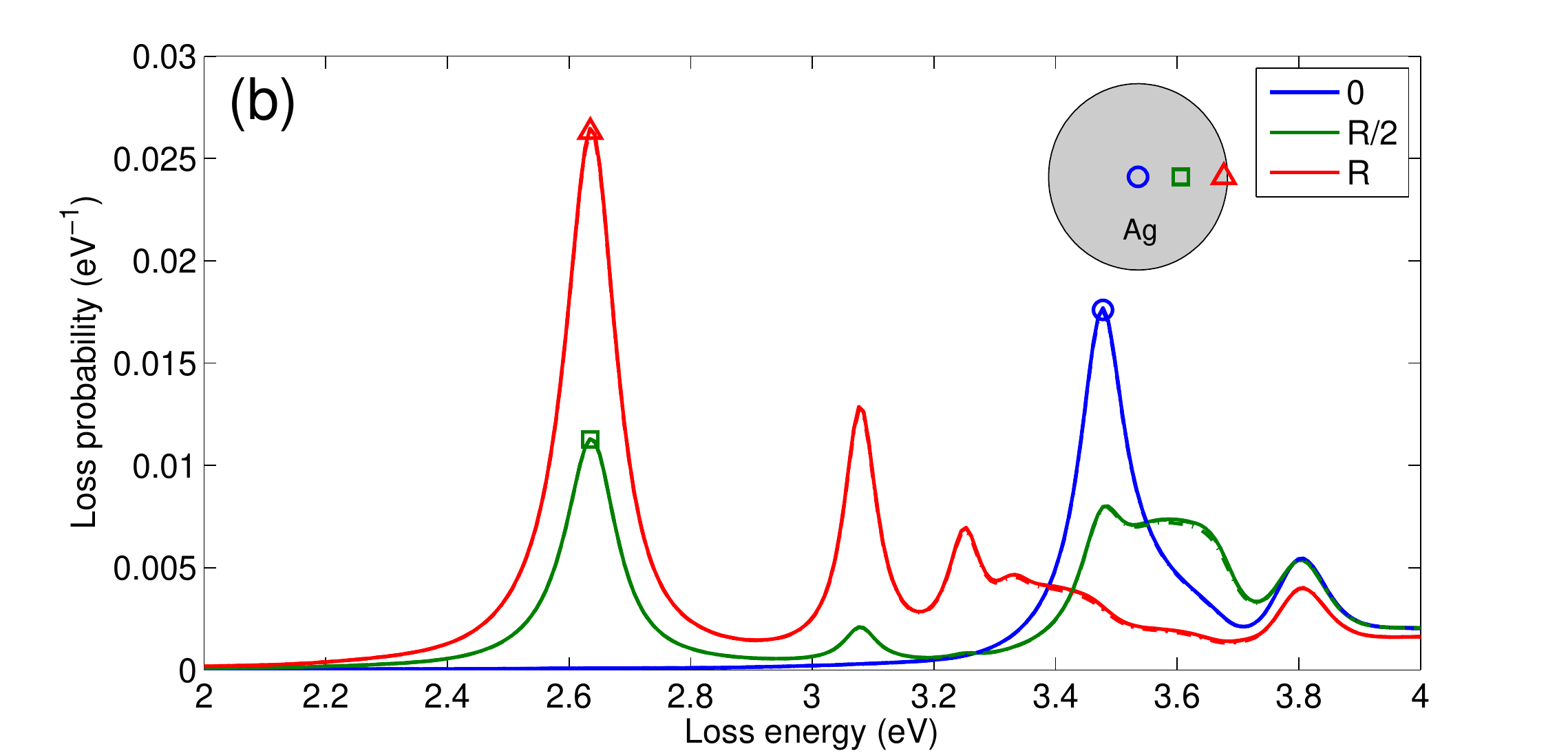}}
\caption{(a) Zoom of the discretized particle boundary for an electron beam passing through the nanoparticle.  The density map (red cross at center) reports the broadened electron distribution, where the broadening is controlled by the \texttt{width} parameter.  \texttt{'cutoff'} determines those boundary elements where the external potentials $\phi_{\rm ext}$ and $\bm A_{\rm ext}$ of Eq.~\eqref{eq:lienard} are integrated over the boundary element.  The boundary element integration is controlled by the \texttt{'refine'} and \texttt{'rule'} properties, as described in more detail in the text.  In the figure we set \texttt{refine=2}. (b) EEL spectra for a silver nanodisk with a diameter of 60 nm and a height of 10 nm.  The impact parameters of the electron beams for the different spectra are reported in the inset, and the beam propagation direction is the $z$-direction perpendicular to the shaded disk.  We investigate different \texttt{width} parameters of 0.1 nm (solid lines), 0.2 nm (dashed lines), and 0.5 nm (dashed-dotted lines), finding practically no differences in the results.  The \texttt{cutoff} parameter is set to 10 nm.}\label{fig:beam}
\end{figure}

To summarize, in the following we list the most important properties of the \texttt{eelsret} class
\begin{code}
exc = eelsret( p, impact, width, vel );    %  initialization
pot = exc( enei );                         %  external potentials, see Eq. (5)
[ psurf, pbulk ] = exc.loss( sig );        %  compute surface and bulk loss probabilities
\end{code}
We emphasize that the functionality of the \texttt{eelsret} class is very similar to that of the \texttt{planewaveret} and \texttt{dipoleret} classes, which account for plane wave and dipole excitations.  The only major difference is that \texttt{eelsret} requires the particle boundaries \texttt{p} of the \texttt{comparticle} object already in the initialization.  This is because upon initialization \texttt{eelsret} computes the crossing points between the particle boundaries and the electron trajectories (if the electron passes by the nanoparticle no crossing points are found), and these crossing points are used in subsequent calculations of the potentials and the loss probabilities to speed up the simulation.

\subsection{Electron beam propagation through nanoparticle and EELS maps}\label{sec:penetrate}

We next investigate the situation where the electron beam passes through the nanoparticle.  The working principle is almost identical to the previous case where the electron passes by the nanoparticle, but the \texttt{width} parameter and the different optional properties have to be set with more care.  In the following, we briefly discuss these quantities in more detail.  
\begin{itemize}

\item[] \texttt{width}.  We have discussed in Sec.~\ref{sec:refine} that the integration of the external potentials over the boundary elements can be facilitated if we use a smoothening parameter $\rho_0$ in the calculation of the external potentials, see Eq.~\eqref{eq:phismooth}.  It is important to stress that the integration could be also performed for $\rho\to 0$ and that the finite $\rho_0$ value only facilitates the computation.  In general, \texttt{width} should be chosen smaller than the average size of the boundary elements, as also shown in Fig.~\ref{fig:beam}(a).

\item[] \texttt{'cutoff'}.  The \verb!cutoff! parameter determines those boundary elements where the external potentials become integrated.  In more detail, we select all boundary elements fully or partially located within a circle with radius \texttt{cutoff} and centered around the impact parameter $\bm R_0=(x_0,y_0)$, as shown in Fig.~\ref{fig:beam}(a) by the red circle.  \texttt{cutoff} should be set such that at least all direct neighbours of the boundary element crossed by the electron beam are included.

\item[] \texttt{'refine'} and \texttt{'rule'}.  The integration over the boundary elements is controlled by \texttt{'refine'} and \texttt{'rule'}.  \texttt{refine} gives the number of integration points within a triangle.  Quadfaces are divided into two triangles.  On default \verb!rule=18! is used (see \verb!doc triangle_unit_set! for details), and we recommend to use this value throughout.  With \texttt{refine} one can split the triangles into subtriangles.  Usually the default value \verb!refine=1! should give sufficiently accurate results.

\end{itemize}

\begin{figure}
\centerline{\includegraphics[width=0.85\columnwidth]{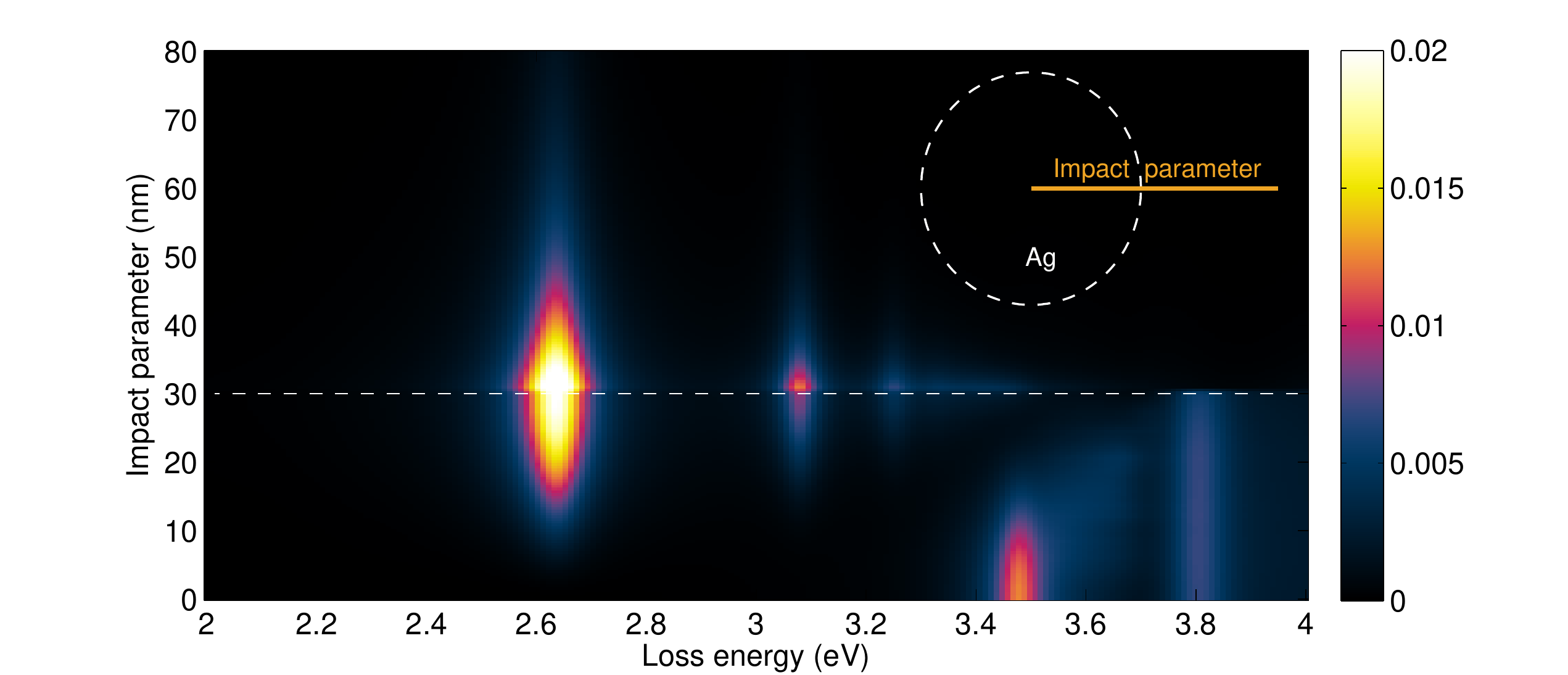}}
\caption{Density map of EEL spectra for nanodisk and for different loss energies and impact parameters, as obtained from the demo program \texttt{demodiskspectrum.m}.  All simulation parameters are identical to those of Fig.~\ref{fig:beam}. }\label{fig:diskmap}
\end{figure}

Figure~\ref{fig:beam}(b) shows the EEL spectra of a nanodisk (for parameters see figure caption) for three impact parameters, which are described in the inset.  In Fig.~\ref{fig:diskmap} we show a density plot of identical loss spectra for a whole range of impact parameters.  One observes a number of peaks, attributed to the dipolar and quadrupolar modes at 2.6 eV and 3.1 eV, respectively, a breathing mode at 3.5 eV~\cite{schmidt:12}, and the bulk losses at 3.8 eV.

\begin{figure}
\centerline{\includegraphics[width=0.85\columnwidth]{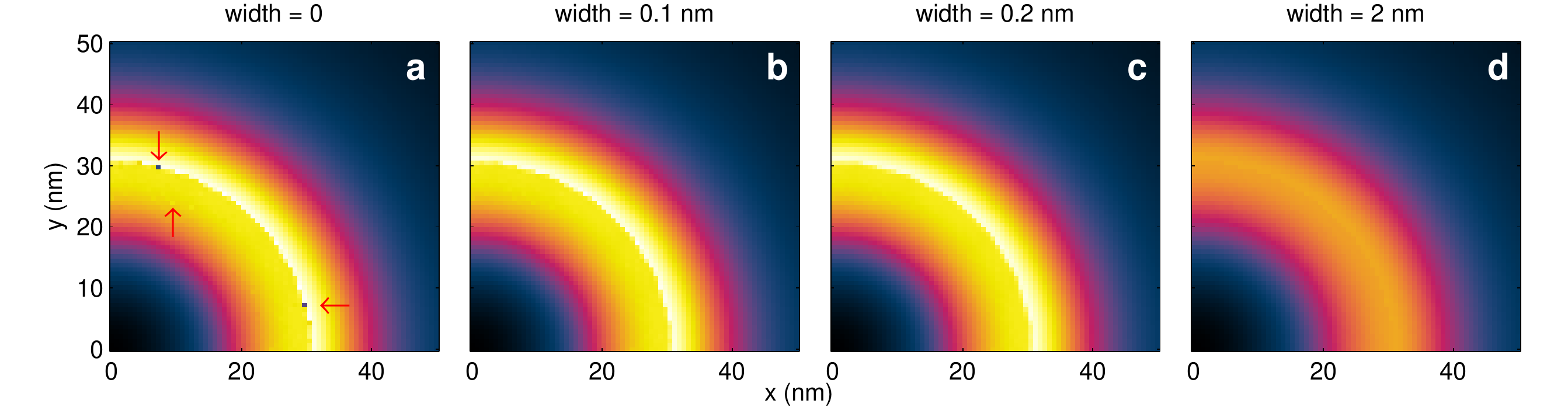}}
\caption{EELS maps for dipolar disk modes and for different \texttt{width} parameters.  We set \texttt{cutoff=10} and use the same simulation parameters as in Fig.~\ref{fig:beam}.  For \texttt{width=0} in panel (a) the computed maps show spikes at certain points, when the impact parameter $\bm R_0$ is too close to a collocation point.  For the moderate \texttt{width} parameters of panels (b,c) the results are sufficiently smooth and almost independent on the chosen value, whereas for too large parameters the EELS map become smeared out, see panel (d).}\label{fig:diskbeam}
\end{figure}

How should one chose the \texttt{width}, \texttt{cutoff}, and \texttt{refine} parameters?  Quite generally, the results depend rather unsensitively on the chosen parameters.  In Fig.~\ref{fig:beam} we show results for various \texttt{width} parameters listed in the figure caption, which are almost indistinguishable.  In Fig.~\ref{fig:diskbeam} we depict EELS maps for the dipolar disk mode at 2.6 eV and for various \texttt{width} parameters.  For \texttt{width=0} in panel (a) one observes for certain impact parameters spikes (some of them indicated with arrows), where the loss probabilities becomes significantly enhanced or reduced in comparison to neighbour points.  This indicates that the impact parameter is located too closely to the collocation point of the boundary element, and the numerical integration fails.  For \texttt{width} parameters of 0.1 or 0.2 nm, panels (b,c), these spikes are absent and the results are almost indistinguishable.  Finally, in panel (d) we report results for a too large smoothening parameter with a significant smearing of the features visible in panels (a--c).  Thus, \texttt{width} should be chosen significantly smaller than the size of the boundary elements but large enough to avoid spikes in the computed EELS maps. 

Let us finally briefly discuss the simulation of EELS maps for a nanotriangle, as computed with the demo program \texttt{demotrianglemap.m}.  Results are shown in Fig.~\ref{fig:trianglemap}.  First we set up an array $\bm R=(x,y)$ of impact parameters and initialize the \texttt{eelsret} object.
\begin{code}
%  mesh for impact parameters
[ x, y ] = meshgrid( linspace( - 70, 50, 50 ), linspace( 0, 50, 35 ) );  
impact = [ x( : ), y( : ) ];           %  impact parameters
vel = eelsbase.ene2vel( 200e3 );       %  electron velocity
[ width, cutoff ] = deal( 0.2, 10 );   %  width of electron beam and cutoff parameter
exc = eelsret( p, impact, width, vel, 'cutoff', cutoff );
\end{code}
Note that in the initialization of \texttt{exc} we pass a matrix \verb![x(:),y(:)]! of impact parameters.  As for the BEM solver, we recommend to use for the boundary element integration the same or larger \verb!'cutoff'! and \verb!'refine'! values as for the \texttt{eelsret} object (see Ref.~\cite{hohenester.cpc:12} and toolbox help pages for further details).
\begin{code}
op = green.options( 'cutoff', 20, 'refine', 2 );   %  options for face integration
bem = bemret( p, [], op );                         %  set up BEM solver
\end{code}
Finally, once the loss probabilities are computed one should reshape \texttt{psurf} and \texttt{pbulk} to the size of the impact parameter mesh.
\begin{code}
p = reshape( psurf + pbulk, size( x ) );  %  reshape loss probability
\end{code}

\subsection{Electric field and cathodoluminescence}\label{sec:efield}

\begin{figure}
\centerline{\includegraphics[width=0.75\columnwidth]{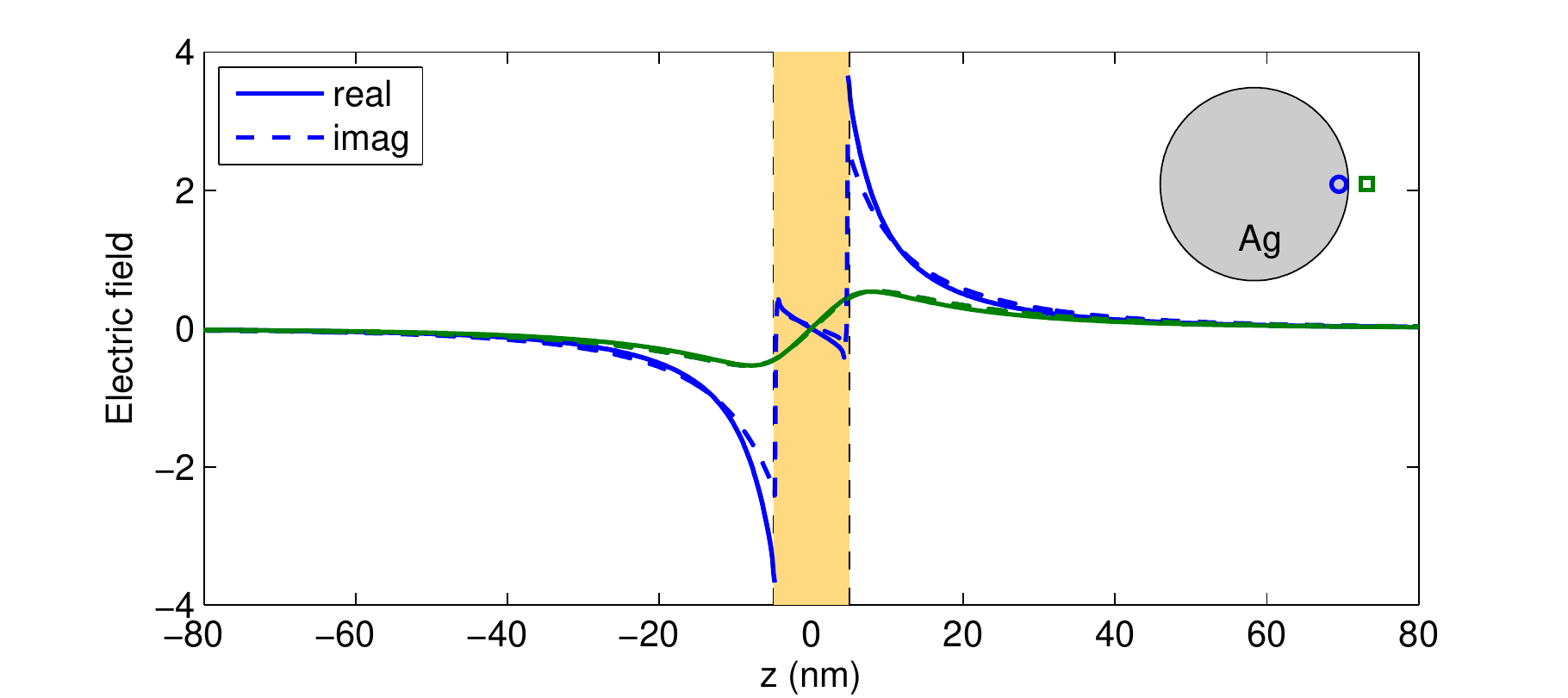}}
\caption{Electric field in $z$-direction for dipolar disk mode.  The simulation and disk parameters are identical to those of Fig.~\ref{fig:beam}, the fields are computed with \texttt{demodiskfield.m}.}\label{fig:diskfield}
\end{figure}

In some cases it is useful to plot the electromagnetic fields induced by the electron beam.  We briefly discuss how this can be done.  A demo program is provided by \texttt{demodiskfield.m}, and the simulation results are shown in Fig.~\ref{fig:diskfield}.

We first set up the \texttt{eelsret} object and the BEM solver, following the prescription given above, and compute the surface charge and current distributions.
\begin{code}
exc = eelsret( p, [ b, 0 ], width, vel, 'cutoff', cutoff );   %  EELS excitation
bem = bemret( p, [], op );                                    %  BEM solver
sig = bem \ exc( enei );                                      %  surface charges and currents
\end{code}
Next, we define the points where the electric field should be computed, using the \texttt{compoint} class of the \texttt{MNPBEM} toolbox, and define a Green function object \texttt{compgreen}.
\begin{code}
z = linspace( - 80, 80, 1001 ) .';                            %  z-values for field
pt = compoint( p, [ b + 0 * z, 0 * z, z ], 'mindist', 0.1 );  %  convert to points
g = compgreen( pt, p, op );                                   %  Green function object

field = g.field( sig );  %  electromagnetic fields at PT positions
e = pt( field.e );       %  extract electric field
\end{code}
In the last two lines we compute the electromagnetic fields, and extract the electric field.  The command \texttt{e=pt(field.e)} brings \texttt{e} to the same form as \texttt{z}, setting fields at points too close to the boundary (which we have discarded in our \texttt{compoint} initialization with the parameter \texttt{'mindist'} \cite{hohenester.cpc:12}) to \texttt{NaN}.  Figure~\ref{fig:diskfield} shows simulation results.  For the electron beam passing through the nanodisk, the field amplitude $E_z$ increases strongly in vicinity of the nanoparticle, which we attribute to evanescent plasmonic fields, and $E_z$ is very small inside the nanodisk because of the efficient free-carrier screening inside conductors.

With the \texttt{MNPBEM} toolbox it is also possible to compute the light emitted from the nanoparticles, the so-called \textit{cathodoluminescence}.  To this end, we first set up a \texttt{spectrumret} object for the calculation of scattering spectra, determine the electromagnetic fields at infinity, and finally compute the scattering spectra.
\begin{code}
spec = spectrumret;               %  set up sphere at infinity
field = farfield( spec, sig );    %  compute far fields at infinity
sca = scattering( spec, field );  %  scattering cross section
\end{code}
In the initialization of \texttt{spec} one could also use a sphere segment rather than the default unit sphere, e.g., to account for finite angle coverages of photodetectors.

\subsection{Parallelization}

Efficient parallelization can be achieved for typical energy loops of the form:
\begin{code}
for ien = 1 : length( enei )                               %  loop over energies
  sig = bem \ exc( enei( ien ) );                          %  compute surface charges
  [ psurf( :, ien ), pbulk( :, ien ) ] = exc.loss( sig );  %  loss probabilities
end
\end{code}
We can replace this loop with:
\begin{code}
matlabpool open;         %  open pool for parallel computation

parfor ien = 1 : length( enei )         %  parallel loop over energies
 sig = bem \ exc( enei( ien ) );        %  same as above  ...
  [ psurf( :, ien ), pbulk( :, ien ) ] = exc.loss( sig ); 
end
\end{code}
The important point is that all computations inside the loop can be performed independently, as is the case for the BEM simulation as well as the calculation of the external potentials and loss probabilities.

\subsection{Quasistatic limit}

\begin{figure}
\centerline{\includegraphics[height=0.35\columnwidth]{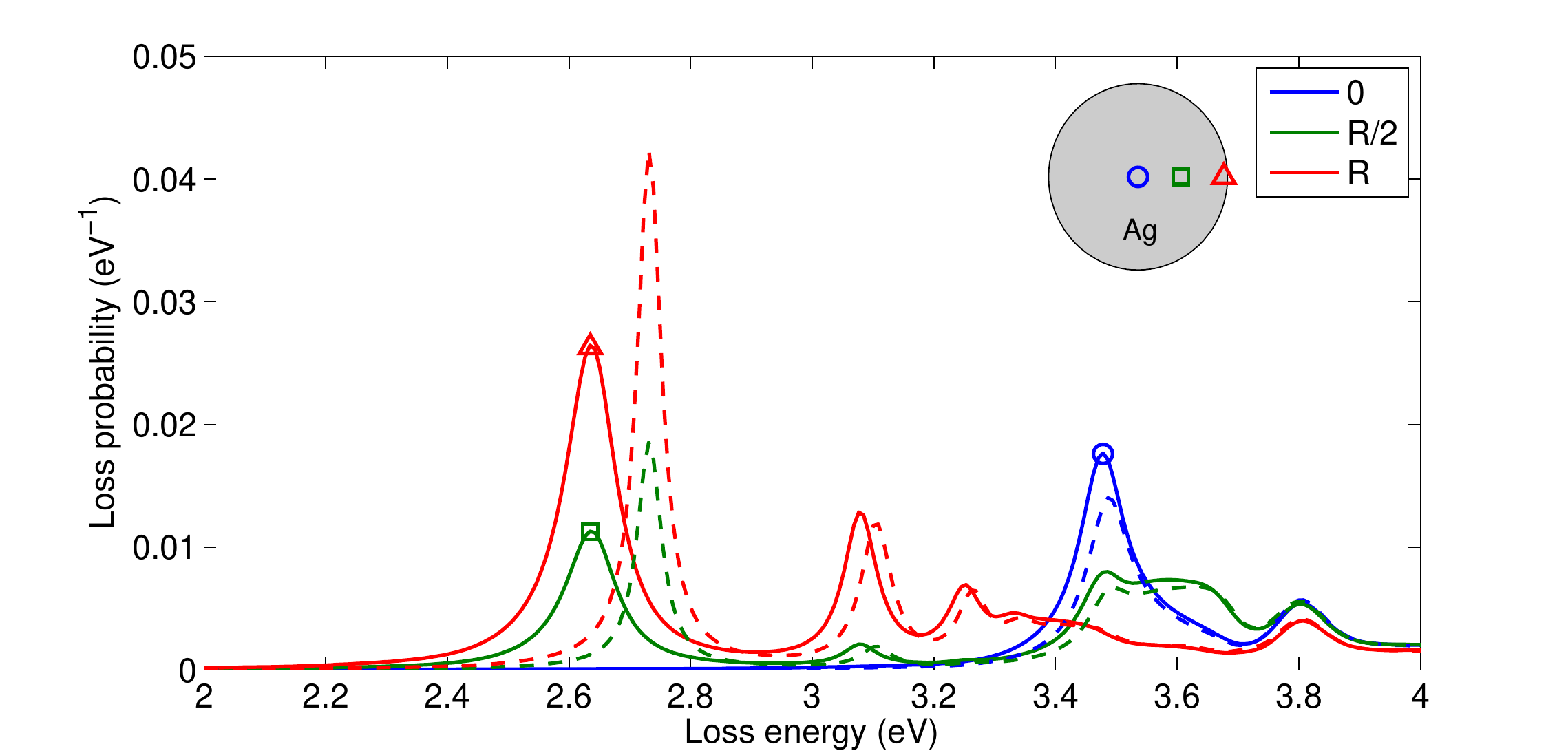}}
\caption{Same as Fig.~\ref{fig:beam}, but computed within the quasistatic limit using the demo program \texttt{demodiskspectrumstat.m}.  The solid lines report simulation results for the full Maxwell equations (same as Fig.~\ref{fig:beam}), the dashed lines report results for the quasistatic approximation.  For the dipolar mode at lowest energy, the peak position and width somewhat differ due to retardation effects and radiation damping.}\label{fig:diskspectrumstat}
\end{figure}

The implementation of the quasistatic limit within the \texttt{eelsstat} class closely follows the retarded case.  The demo program \texttt{demodiskspectrumstat.m} is very similar to \texttt{demodiskspectrum.m} discussed in Sec.~\ref{sec:penetrate}.  We first set up a disk-like nanoparticle and specify the electron beam parameters.  Next, we initialize an \texttt{eelsstat} object by calling
\begin{code}
exc = eelsstat( p, b, width, vel, 'cutoff', cutoff, 'refine', 2 );   %  EELS excitation
\end{code}
The definition of the various parameters is identical to the retarded case.  Finally, we set up the quasistatic \texttt{bemstat} or \texttt{bemstateig} BEM solver~\cite{hohenester.cpc:12}, and compute the surface charge distribution and the energy loss probability using the equations presented in Sec.~\ref{sec:quasistatic}
\begin{code}
bem = bemstat( p, [], op );        %  set up quasistatic BEM solver

for ien = 1 : length( enei )       %  loop over energies
  sig = bem \ exc( enei( ien ) );  %  compute surface charge
  [ psurf( :, ien ), pbulk( :, ien ) ] = exc.loss( sig );   %  electron energy loss probability                 
end
\end{code}
Simulation results are shown in Fig.~\ref{fig:diskspectrumstat}.  We observe that the results of the full and quasistatic simulations are very similar, and only for the dipolar mode at lowest energy the peak position and width somewhat differ due to retardation effects and radiation damping

\section*{Acknowledgment}

I am grateful to Andi Tr\"ugler for most helpful discussions, and thank him as well as Toni H\"orl, J\"urgen Waxenegger, and Harald Ditlbacher for numerous feedback on the simulation program.  This work has been supported by the Austrian Science Fund FWF under project P24511--N26 and the SFB NextLite.



\bigskip


\end{document}